\documentclass[]{aastex631}
\usepackage{amsmath}

\newcommand{\eb}{\begin{equation}}
\newcommand{\ee}{\end{equation}}

\newcommand{\uas}{$\mu$as}

\newcommand{\phe}{\texttt{phot\_bp\_rp\_excess\_factor}}
\newcommand{\gof}{\texttt{astrometric\_gof\_al}}

\definecolor{rkka}{RGB}{219,66,32}
\definecolor{nsgreen}{rgb}{0.1,0.5,0.1}

\shorttitle{Radio-Optical Reference Catalog}
\shortauthors{Makarov et al.}

\begin{document}

\title{Radio-Optical Reference Catalog, version 1}

\correspondingauthor{Valeri V. Makarov}
\email{valeri.makarov@gmail.com}

\author[0000-0003-2336-7887]{Valeri V. Makarov}
\affiliation{U.S. Naval Observatory, 3450 Massachusetts Ave NW, Washington, DC 20392-5420, USA}

\author[0000-0002-4146-1618]{Megan C. Johnson}
\affiliation{U.S. Naval Observatory, 3450 Massachusetts Ave NW, Washington, DC 20392-5420, USA}

\author[0000-0002-4902-8077]{Nathan J. Secrest}
\affiliation{U.S. Naval Observatory, 3450 Massachusetts Ave NW, Washington, DC 20392-5420, USA}

\begin{abstract}
The fundamental celestial reference frame (CRF) is based on two catalogs of astrometric positions, the third realization of the International Celestial Reference Frame (ICRF3), and the much larger Gaia~CRF, built from the third data release (DR3). The objects in common between these two catalogs are mostly distant AGNs and quasars that are both sufficiently optically bright for Gaia and radio-loud for the VLBI. This limited collection of reference objects is crucially important for the mutual alignment of the two CRFs and maintenance of all the other frames and coordinate systems branching from the ICRF. In this paper, we show that the three components of ICRF3 (S/X, K, and X/Ka band catalogs) have significantly different
sky-correlated vector fields of position offsets with respect to Gaia~DR3. When iteratively expanded in the vector spherical harmonics up to degree 4 on a carefully vetted set of common sources, each of these components includes several statistically significant terms. The median sky-correlated offsets from the Gaia positions are found to be 56 \uas\ for the S/X, 100 \uas\ for the K, and 324 \uas\ for the Ka catalogs. The weighted mean vector field is subtracted from the Gaia reference positions, while the deviations from that field are added to each of the ICRF3 components. The  corrected positions from each of the four input catalogs are combined into a single weighted mean catalog, which we propose to be the current most accurate realization of an inertial radio-optical CRF.
\end{abstract}

\section{Introduction} \label{section: Introduction}

The succession of International Celestial Reference Frame (ICRF) catalogs \citep{1998AJ....116..516M, 2009ITN....35....1M, 2015AJ....150...58F, 2020A&A...644A.159C} established an evolving gold standard for a non-rotating and stable coordinate system defined by the position of sources on the sky. This endeavor capitalizes on the gradually increasing precision, stability, and temporal baseline of astrometric measurements using very long baseline interferometry (VLBI) in four radio passbands (the geodetic S/X bands, and K/Ka), as well as the favorable properties of quasars and distant AGNs, which are assumed to have negligible proper motions and compact appearances. The hegemony of radio astrometry in fundamental astronomy has been diminished by the advent of the Gaia space astrometry mission \citep{2016A&A...595A...1G}, which has achieved comparable astrometric precision at visual (optical) wavelengths for a much larger number of sources \citep{2021A&A...649A...1G}. The resultant Gaia Celestial Reference Frame \citep[Gaia-CRF3, hereafter GCRF;][]{2022A&A...667A.148G} therefore rivals the ICRF, while not being independent of it because the overall orientation of the axes of GCRF is as close to ICRF3 as possible for the sake of consistency and practical convenience \citep{2018A&A...616A..14G}. The proper motion system of GCRF, on the other hand, is independently anchored to the frame motion defined by distant quasars, which is supposed to be negligibly small. However, there may nonetheless be large-scale, position-correlated or systematic differences between the three ICRF3 frames (S/X, X/Ka, and K) and the GCRF, as traced by their shared sources. Determining the sky-correlated differences between the three components of ICRF3 and Gaia-CRF is the only direct way to link the latter to ICRS.

The main objective of this paper is to accurately determine such systematic differences and remove these discrepancies as much as possible to create a merged astrometric catalog that is maximally consistent at both optical and radio wavelengths. This catalog, which we call the Radio-Optical Reference Catalog (RORC), truly represents the state of the art in the quest for the most accurate, stable, and inertial reference frame, and is the first of a succession of such catalogs that will be created as more accurate astrometric data become available. In Section~\ref{section: Methodology}, we describe how the set of sources shared between ICRF3 and GCRF were selected, paying mind to systematic issues with these catalogs that have become known in recent years. In Section~\ref{vsh.sec}, we describe the vector spherical harmonic (VSH) method we use to analyze sky-correlated systematics, and provide the VSH fits for position offsets of Gaia DR3 with respect to each of the ICRF3 constituents. Section \ref{pca.sec} is devoted to the important problem of robustness of the determined systematic fields, which is expressed via the spectra of singular values. We discuss these findings in Section~\ref{disc.sec} and develop a general approach to combining the four reference catalogs into a single radio-optical CRF. The final construction of the first version of RORC, RORC-1 is detailed in Section~\ref{con.sec}, and we provide some concluding remarks in Section~\ref{end.sec}.

\section{Methodology} \label{section: Methodology}
\subsection{Near neighbor analysis and cross-match radius}
As described in \citet{2020A&A...644A.159C}, the ICRF3 catalog consists of three frames, which partly overlap in sources measured with VLBI networks at four wavelengths, including the standard geodetic X-band (8.4~GHz), the S-band (2.3~GHz) used only for calibration purposes,
the K~band (24~GHz), and the Ka~band (32~GHz). The combinations of simultaneous measurements in the observing and calibration frequencies resulted in the separate S/X, the K, and the X/Ka catalogs. We crossmatched the S/X catalog with the Gaia DR3 main catalog \citep{2016A&A...595A...1G, 2021A&A...649A...1G} directly using the mean positions and a search radius of $11\arcsec$. The wide search radius was meant to help identifying potentially problematic sources in crowded areas and confusion sources, because some of the radio-loud quasars do not have detectable optical counterparts, especially at low Galactic latitudes. Double quasars and close optical pairs with field stars generate significant astrometric perturbations including bogus proper motions or gross errors that are not captured by the formal uncertainties \citep{2017ApJ...840L...1M, 2022ApJ...933...28M} in the Gaia data. The impact of chance and genuine neighbors depends on the relative brightnesses and angular separations of sources. The most dangerous companions are those closer than $\sim2\arcsec$ \citep[where Gaia EDR3/DR3 loses source completeness;][]{2021A&A...649A...5F}, while more distant ones do not seem to astrometrically perturb our targets unless they are much brighter. We investigated in depth several sources from the S/X sample in the most crowded areas and could not find any obvious relation of ICRF$-$Gaia position differences to chance neighbors outside of $2\arcsec$. Therefore, a smaller cross-match radius of $2\arcsec$ was used for the K and Ka samples.

Of the 4536 sources in the S/X catalog, 3986 have at least one association in Gaia DR3 within $11\arcsec$. 
Hence, at least 550 radio sources do not have any optical Gaia associations. The number of invisible ICRF sources is higher in fact, as is seen in the example of the source ICRF~J095551.5+694045, which has 52 optical associations. Furthermore, we find 3503 (out of 8404) matches with separations below $0\farcs9$. All of the 3987 S/X associations within $11\arcsec$ have total positional errors less than 119~mas, so there is little reason to expect a true counterpart at five to seven times this value. Such sources cannot be included in RORC. We therefore set a much stricter limit on the separation between the radio source and the closest Gaia association of $0\farcs5$. After this cut, the total number of the remaining matches is 3503. This is only an intermediate cut, however, which is superseded by the following more stringent criteria, resulting in a sample with a largest position offset of 14~mas.

\subsection{Selecting the precious sets}
\label{sel.sec}
Shortly after the first release of Gaia mission data, it became obvious that a large fraction of ICRF sources have discrepant positions in the radio and optical domains at the level of several combined formal errors and higher \citep{2017ApJ...835L..30M, 2017MNRAS.471.3775P}.  The subsequent data releases from the Gaia mission with much improved astrometric precision of fainter sources only confirmed and strengthened this discrepancy \citep{2019ApJ...873..132M, 2019MNRAS.482.3023P}.
In nearby galaxies, the offsets are larger and easier to detect, showing a range of possible causes, including the presence of dust lanes and asymmetries affecting the optical centroiding \citep{2018MNRAS.475.5179S}. Caution is needed when interpreting Gaia astrometry in extended objects \citep[e.g.,][]{2022A&A...660A..16S, 2022ApJ...933...28M}, however. Nonetheless, a significant fraction of AGNs listed in the ICRF3 S/X catalog (including defining radio sources) reside in nearby galaxies \citep{2012MmSAI..83..952M}. A lower bound on this fraction can be indirectly determined by using the distribution of the \phe\ parameter and the fact that the vast majority of nearby galaxies have this value above 2 \citep[][their Fig. 2]{2022ApJ...933...28M}. This gives an estimate of $>11$\% for the S/X catalog, for example. The majority of sources at higher redshifts may be astrometrically perturbed on the ICRF3 side by radio-emitting, possibly temporally unstable, jets offset from the central cores. More rare occurrences include unresolved dual AGNs, alignments with foreground stars, and gravitationally lensed images.

\begin{figure*}
    \includegraphics[width=0.47 \textwidth]{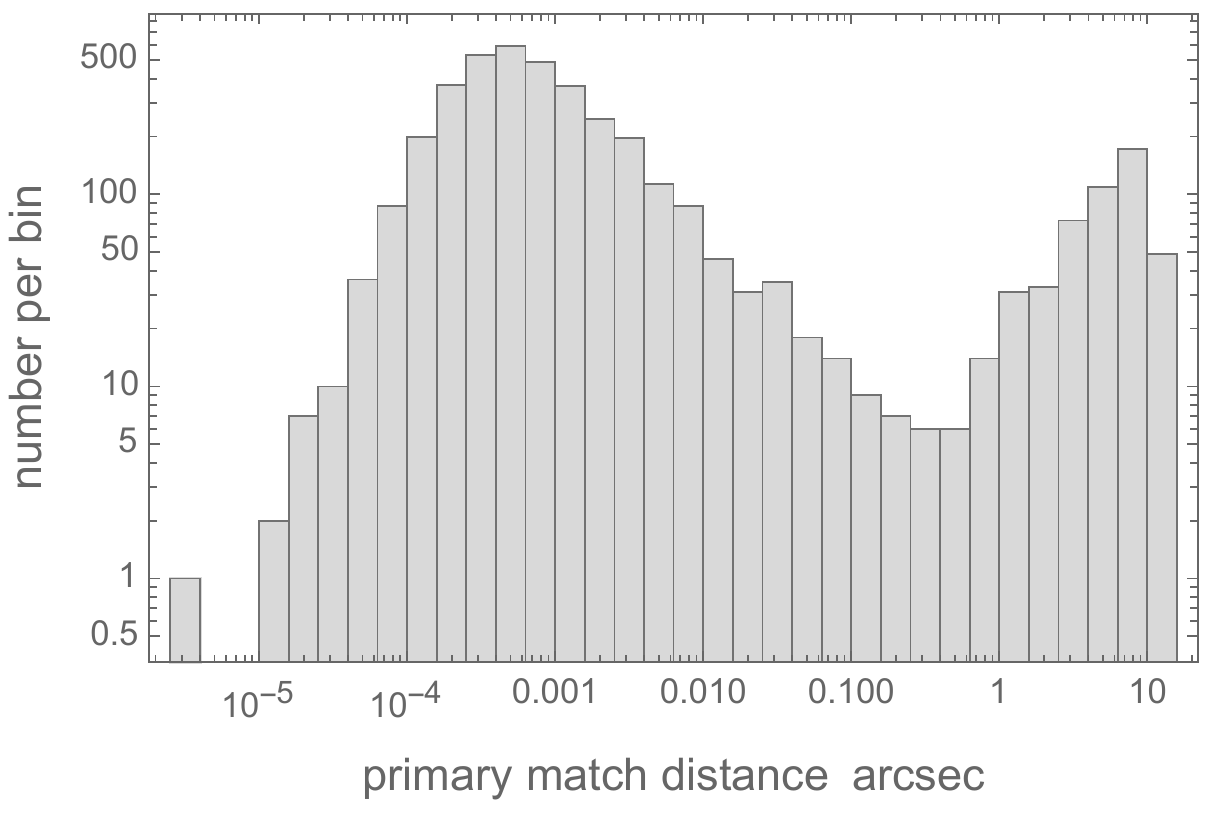}
    \includegraphics[width=0.47 \textwidth]{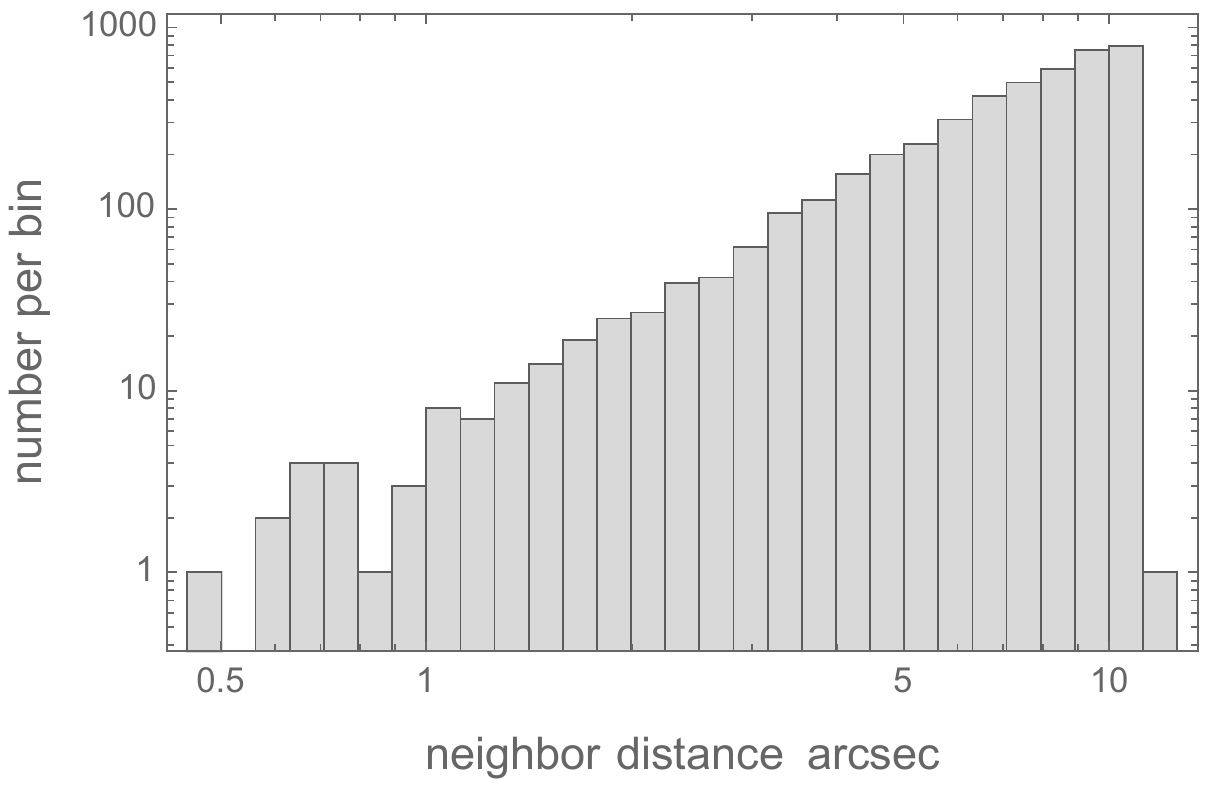}
    \caption{Histograms of angular separations of ICRF3~S/X sources from matched Gaia DR3 counterparts in the log-log scale.
    Left: separations of the primary, i.e., nearest matches; Right: separations of the second nearest neighbors, if any.}
    \label{sxsep.fig}
\end{figure*}

Figure~\ref{sxsep.fig} shows the sample distributions of angular separations between the target positions from the ICRF3 S/X sample and their nearest (left panel) and second-nearest (right panel) optical matches, which are called primary matches and neighbors, respectively. The primary matches clearly obey a bimodal distribution, where the larger peak is composed of true optical counterparts of the radio sources. The secondary peak in the left graph represents mostly optical pairs and double quasars, to be discarded. The neighbors in the right graph, on the other hand, show a linearly rising distribution with distance, as expected of random interlopers in the log-log scale. We note that a small excess of neighbors is present at separations below $1\farcs5$ where the source completeness of Gaia~DR3 is degraded \citep{2021A&A...649A...5F} and we should expect to see a dip. These tightly separated 44 neighbors represent surviving genuine dual AGNs and microlensed multiple images. For example, the tightly packed optical images of the source ICRF J191414.5+012426 suggest an unresolved lens, while some more spread out components of possible lenses have already been noted in the literature \citep{2019A&A...622A.165D}, such as ICRF J014737.7+485937 and ICRF J174037.1+031147. In all these special cases, the true optical counterpart is much closer to the target's radio position. This observation allows us to limit our analysis to the 3503 primary matches.

\begin{figure*}
    \includegraphics[width=0.48 \textwidth]{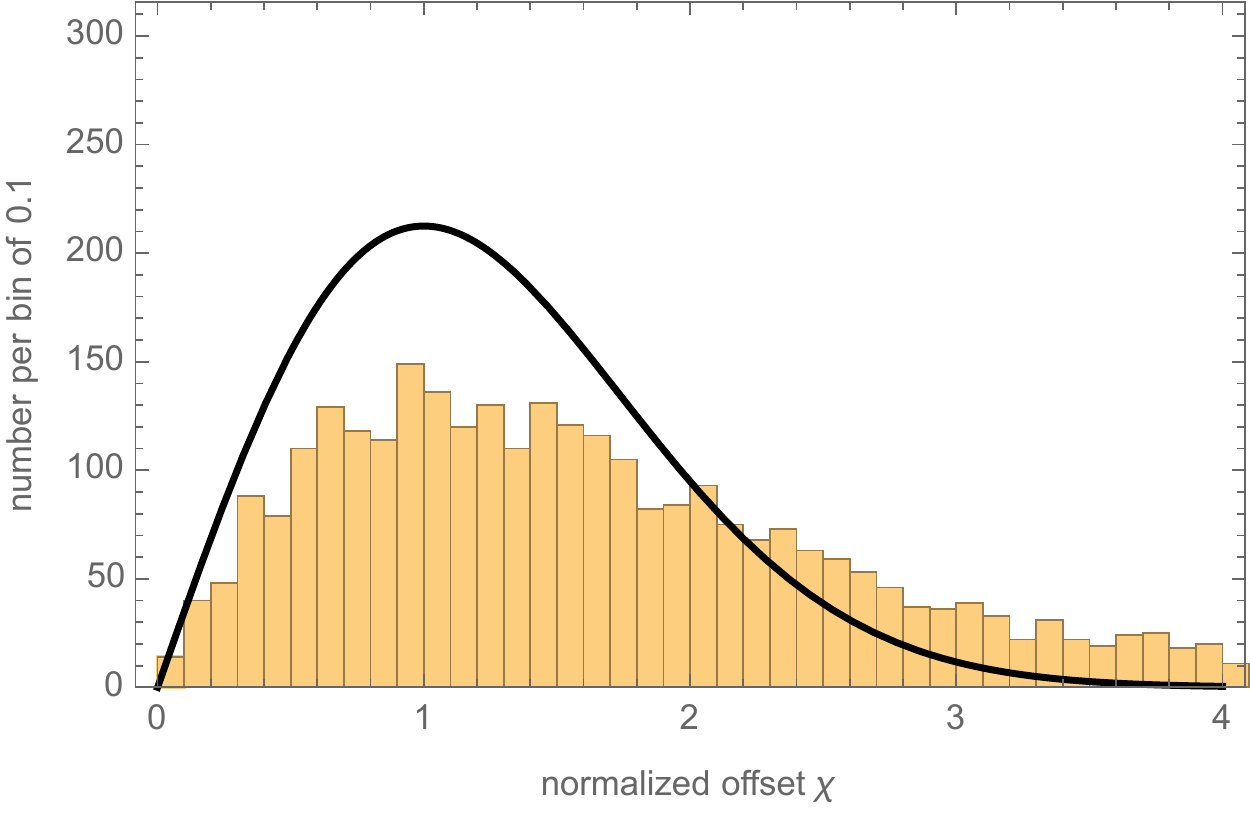}
    \includegraphics[width=0.48 \textwidth]{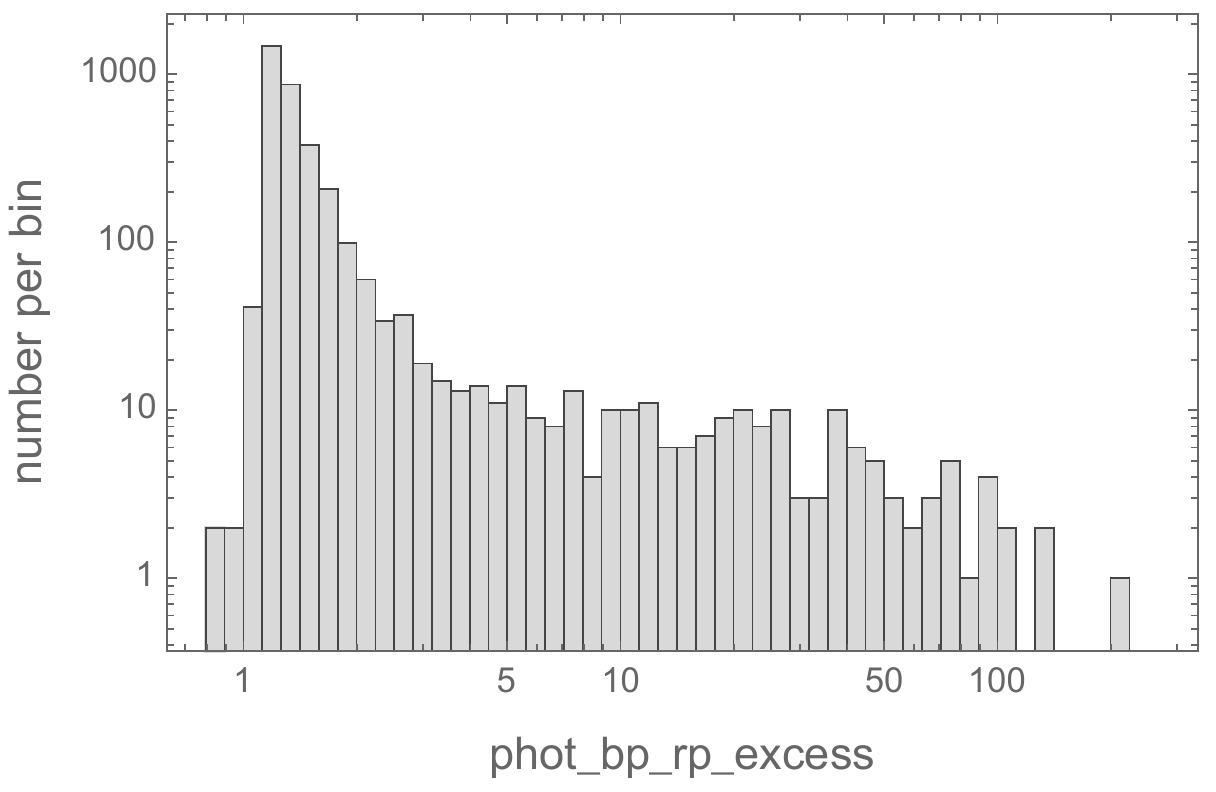}
    \caption{Left: Histogram of the $X$-values of the radio-optical position offsets of S/X sources matched in ICRF3 and Gaia DR3. The expected Rayleigh $(\nu=1)$ distribution is shown with the black line. Right: log-log histogram of the \phe\ values for the same sources from Gaia DR3.}
    \label{sxray.fig}
\end{figure*}

For each pair of radio and optical sources, the tangential offset vector and its normalized length $X$  is computed from the available positions and their 2$\times2$ covariance matrices as described in \citep{2017ApJ...835L..30M,2018A&A...616A..14G}.\footnote{We adopt the capital $X$ notation from \citet{2016A&A...595A...5M}.} These statistical values are expected to follow a Rayleigh distribution with $\nu=1$ if the measurement errors are Gaussian, the formal errors correctly represent the actual dispersion, and there are no intrinsic position offsets between the ICRF sources and their Gaia counterparts. This distribution is shown in the left panel of Figure~\ref{sxray.fig}. The scatter of observed offsets is profoundly larger than the formal dispersion estimates, which is seen in the deficit of values around the peak value of 1 and an excess of values in the tail of the Rayleigh distribution above 2. For the S/X sample, specifically, we find a rate of 32\% of objects with $X>2.5$, 25\% for $X>3.0$, 21\% for $X>3.5$, and 18\% for $X>4.0$. 

A number of factors that can perturb the Gaia astrometric measurements of quasars and AGNs have been identified in the literature \citep[e.g.,][]{2012MmSAI..83..952M}. Dual quasars and optical pairs with field stars are associated with detectably elevated measured proper motions \citep{2022A&A...660A..16S,2022ApJ...933...28M}. However, the main perturbation of Gaia astrometric and photometric measurements for the S/X sample comes from the extended character of host galaxies at low redshift. The Gaia DR3 data processing pipeline is not adjusted for treating extended images, which, in the case of nearby AGN-harboring galaxies, are often asymmetric or misaligned due to the obscuring dust structures and merger morphology. The astrometric degradation caused by the extended structures at $z<0.5$ is strongly correlated with the photometric excess parameter \phe\ \citep{2022ApJ...933...28M}. Spectroscopic redshifts are only known  for a fraction of the S/X sample \citep[1014/4536 = 22\%, ][]{2022ApJS..260...33S}, so we rely in this study on the \phe\ parameter to clean our working sample of the perturbed nearby sources. The set of requirements used in this study to select the precious set of most stable and unproblematic radio-optical reference sources is: 1) $X<2.5$; 2) \phe$<4$; 3) \texttt{astrometric\_gof\_al}$<5$; 4) \texttt{ruwe}$<2$. The first two filters are responsible for the greatest reduction of the working sample, with the last two removing only a few objects. Filter 1 leaves only objects that distributed mostly within the expected curve in Figure~\ref{sxray.fig}. 
Selecting such a low threshold value for $X$ may seem overly aggressive, but the choice is justified by the statistics of position outliers. At $X<2.5$, 1107 sources are removed from the crossmatched S/X sample. Relaxing this filter to $X<3.0$ eliminates 876 sources. From the Rayleigh(1) PDF distribution expected for the normalized position offsets, the number of purely stochastic outliers is 154 at $X>2.5$ and 39 at $X>3.0$. Hence, using a cut at 3 instead of 2.5 preserves 115 stochastic outliers and removes 116 genuinely perturbed sources. In the context of this work, it is important to remove perturbed data with extra noise not captured by the formal covariances, because the latter are used for weighting the condition equations of the VSH fit (Section \ref{vsh.sec}). Due to the large dynamic range of the formal uncertainties and associated weights, the VSH fits are sensitive to even relatively small admixtures of noisy data with unknown statistical properties. Including a hundred more outliers would make the formal uncertainties of the VSH coefficients slightly smaller, at the same time rendering the VSH fit less reliable. This consideration assumes that the genuine perturbations of positions are random in direction, while the main objective of the VSH fitting is to detect sky-correlated  vector fields. This contradiction is resolved by iterating the entire source selection and VSH fitting procedure for each sample as described in Section \ref{iter.sec}.

The distribution of the \phe\ values (not shown for brevity) does not have a clear dip separating the less perturbed and more perturbed sources. We inspected the available optical images of randomly selected Gaia CRF sources with \phe\ above 2 and 4 and determined that most of the images with the latter threshold clearly show the presence of extended structures around the central core even when the entries do not appear to be astrometrically perturbed.  Filters 3 and 4 utilize two parameters as given in the Gaia archival database, which are related to the consistency of the astrometric data with the object-specific model. The number of S/X sources selected after these initial cuts is 2136. 

We use the same method of cross-matching and initial sample generation for the K and Ka components of this study. The K catalog of ICRF3 is composed of 824 sources. The total number of Gaia associations within $11\arcsec$ of the K catalog positions is 1868, but only 785 unique objects have at least one association, so many of them have more than one. The largest number of Gaia associations for a single K~catalog object is 33, none of which are valid matches. Ten radio sources have more than 20 optical matches, including one defining source ICRF~J172155.9$-$255840. This example shows that some of the ICRF sources are not well suited for optical astrometry as they reside in crowded fields. The histogram of the angular distances to the nearest match shows a clear gap around $0\farcs1$. The sources with the second nearest match within $1\farcs5$ (8 in total) include  M87, where, apart from the central core image, Gaia resolved two knots in the spectacular jet of this AGN. 

The histogram of normalized position offsets $X$ for the K sample is similar to that of the S/X sample except for a more prominent Poisson noise component due to the smaller size. 
We find a rate of 25\% of objects with $X>2.5$, 19\% for $X>3.0$, 15\% for $X>3.5$, and 13\% for $X>4.0$. The quantiles of $X$ are consistently smaller than those for the S/X sample by several percent. This, however, should not be interpreted as smaller radio-optical offsets in the K band. The likely reason for the apparently smaller dispersion is the systematically larger formal uncertainty of the K-band astrometry. Indeed, while a systematic noise floor of 30\,$\mu$as was added in quadrature to the formal errors of both R.A.\ and decl.\ for S/X, the decl.\ noise floor of K was set to 50\,$\mu$as \citep{2020A&A...644A.159C}. The formal errors per coordinate peak at about 70--80 \uas\ but have a significant tail stretching to 1 mas and beyond. However, the distribution of \phe\ values is much more compact than the S/X histogram with only 12 sources exceeding 4. Apparently, the target list for these VLBI measurements was more carefully vetted avoiding numerous nearby galaxies on the S/X list. It is also possible that since selection on S/X is more sensitive to steep-spectrum source than selection on K, a higher prevalence of nearby radio galaxies is expected in the former, whereas more flat-spectrum distant blazars might be expected on K. For the initial K working sample, we apply the same set of four filters as described in the previous paragraph. Note that we do not attempt to keep the same objects in the three working samples giving preferences to the objects in common. Inevitably, some of the common objects may be accepted in one working sample but rejected in another due to the natural random error dispersion. As we will see from this study, the systematic part of the radio-optical offsets is significant and unique to a given set, which also precludes a clear separation of ``good" and ``bad" reference sources across the VLBI radio bands. The initial K sample is a collection of 492 carefully matched and astrometrically stable radio-optical sources.

The starting number of sources in the Ka sample is 678. Only 619 of them have any optical counterparts within $2\arcsec$ of the radio positions, the remaining 59 being optically dim or invisible. Only one object, the blazar ICRF~J130817.3$-$670705 situated in a crowded region, has three optical counterparts within $2\arcsec$. Eight Ka sources have optical companions within $1\farcs5$, which are likely to degrade the Gaia astrometric quality, but not necessarily the quality of the radio determinations. For example, the source ICRF~J101544.0+122707 has a bright neighbor separated by $1\arcsec$, which obviously perturbs the Gaia proper motion, while the radio source has no detectable structure from closure phase measurements \citep{2019ApJS..242....5X}.

The distribution of normalized radio-optical position offsets for the Ka sample is significantly worse than those for the S/X and K sets. The histogram is quite flat and a large fraction of the sample spills over the expected Rayleigh distribution. The formal coordinate errors, however, are similar to those in the larger catalogs, with most common values below 100~\uas, and the error floor is the same as S/X \citep{2020A&A...644A.159C}. The problem appears to be on the radio side, because the distributions of \gof\ and \texttt{ruwe} parameters for the primary matches do not show any signs of perturbation. In fact, the selection of sources is markedly better than in the S/X sample from the optical astrometry side. The distribution of \phe\ is also much more compact with only five objects outside our chosen limit 4. Applying the same set of four filters as for the other ICRF components leaves 307 sources. We note that the filter $X <2.5$ is by far the most restrictive in this case discarding almost half of the initial sample.

\section{Vector spherical harmonic fit of Gaia--ICRF position differences}
\label{vsh.sec}

The vector spherical harmonic decomposition technique is a powerful method for analysis of sky-correlated components in position and proper motion fields of astrometric catalogs \citep{2004ASPC..316..230V, 2007AJ....134..367M, 2011AstL...37..874V, 2012A&A...547A..59M, 2021MNRAS.506.5540M}. In these applications, the 3D VSH functions are reduced to their tangential components defined on the unit sphere. The astrometric data fields (position differences or proper motions) are real-valued tangential vectors, and the transition from the complex-valued VSH is achieved by including the real and imaginary parts of the VSH into the fitting set. The mathematical background and technical implementation are described in detail in \citet{2022AJ....164..157M}. The limiting degree of the decomposition defines the scale of distortion patterns that can be captured. For this study, we employed a set of 48 VSH terms to degree 4. The upper degree is limited by the number of data vectors and their distribution on the sphere. Using VSH terms of high degree may result in overfitting, when the normal matrix becomes ill-conditioned because of the gaps in the sky coverage. The wide avoidance zone along the Galactic plane and the lower density of ICRF objects in the southern hemisphere pose the greatest difficulty for this study. 

The input data vectors are the tangential position differences Gaia$-$ICRF computed for each source in the three filtered samples. Each vector is defined in the local tangential coordinate frame with axes in the east and north direction in the equatorial ICRS. The origin of the local frame is at the ICRF position. The VSH functions are of two kinds, called here {\bf MVSH} (magnetic) and {\bf EVSH} (electric). The degree 1 VSH terms are of special significance for astrometric vector fields. The first three {\bf MVSH} terms represent the rigid rotation of the entire frame around the three axes of the equatorial system. The first three {\bf EVSH} terms represent polar glides, which may emerge in the data due to the Galactocentric acceleration of the solar system and the drift of the secular aberration dipole, as discussed in Section \ref{disc.sec}. The higher degree terms represent more complicated patterns on the sky. 

The system of condition equations is solved by the weighted least-squares method. We note that the right-hand side of the system includes 2-vectors instead of the more familiar scalars. Consequently, the design matrix has the dimensions $N_{\rm obs} \times 48\times 2$. The weight of each data vector is a $2\times 2$ matrix, which is computed from the sum of the two formal covariance matrices. Blockwise premultiplication of both sides of the equation folds in the dimension spanned by the data vectors and leads to a compact $48\times 48$ normal matrix. The inverse of this matrix provides the full covariance matrix of the solution vector composed of 48 VSH coefficients. The diagonal elements of the solution vector covariance provide the formal dispersions of the estimated VSH coefficients. The numerical values of the coefficients are in the same units of angular  distance as the data vectors (e.g., mas or \uas), but they are not equivalent to the magnitude of the corresponding VSH term because of the normalization coefficients. Each VSH term can be expressed via trigonometric functions of the celestial coordinates. 

\subsection{Iterations of selection and VSH fitting}
\label{iter.sec}
The working sample of unperturbed samples, and, ultimately, the precious set for each ICRF3 component is mostly defined by the most effective cut $X<2.5$, as described in Section \ref{sel.sec}. This filter is responsible for elimination of at least a quarter of the source catalog. The observed position offsets $\boldsymbol{r}_{\rm Gaia}-\boldsymbol{r}_{\rm ICRF}$ are 3D vectors, which are first projected onto the tangential plane for each source, resulting in 2D offset vectors in angular units. These are further normalized using the combined formal covariance matrices to yield the (dimensionless) normalized offset magnitudes $X$. Since the observed position differences are involved in the computation of $X$, placing an upper limit on the latter is equivalent to employing a Bayesian prior that the observed offsets are purely stochastic vectors with a first moment at zero, for each source. This is logically inconsistent with our goal of detecting significant systematic or sky-correlated component of the observed field expressed via the VSH functions. Therefore, a single selection procedure followed by the VSH fit should result in a vector field biased toward zero. This happens because sources located in areas of the sky with significant systematic offsets tend to be rejected more frequently as the probability of the normalized offset $X$ to exceed the adopted limit is higher for them.

To break this {\it circulus vitiosus}, we perform iterations of the entire selection and VSH fitting algorithm. Each iteration starts with computing the previously fitted vector field for a given point $\{\alpha, \delta\}$:
\eb
\boldsymbol{d}(\alpha,\,\delta)=
\sum_{j=1}^{48} a_j\;{\bf VSH}(\alpha,\,\delta),
\label{fit.eq}
\ee 
where {\bf VSH} inludes both {\bf MVSH} and {\bf EVSH} vector-valued functions, $a_j$ are the fitted VSH coefficients, and we simplified the indexing of the terms employing a single running index $j$, which counts the terms of all degrees, orders, and kinds. For the starting iteration all the $a_j$ coefficients are set to zero. This vector field is subtracted from the observed position offsets for each source in the initial catalog (prior to filtering). The normalized position offsets $X$ are then recomputed using the same formal covariances. The subsequent selection of the precious set employs the same set of criteria (Sect. \ref{sel.sec}) but results in a different sample because of the updated $X$. The weighted VSH fit is computed on the updated data set, and the coefficients $a_j$ are filed for the next iteration.

This process converged for all three ICRF3 parts within several iterations. The convergence manifests itself in: 1) the number of selected sources stops growing and becomes fixed or slightly oscillates; 2) the absolute values of the VSH coefficients $a_j$ of the most significant terms begins to oscillate around fixed values; 3) the absolute robustness of the VSH solution saturates at a higher value. We performed different numbers of iterations for the input samples: four for S/X, three for K, and five for Ka. The fastest convergence was achieved for the K sample, which also showed relatively small updates of the VSH fits. The most substantial adjustment of the selected sample and the number of statistically significant VSH terms was seen for the larger S/X sample. The number of accepted sources increased by 32, and the number of VSH terms with S/N above 3 increased from 1 to 7. The iterated VSH fit showed more structure and finer detail than the initial iteration. The median length of the fitted vectors also increased from 33 \uas\ to 56 \uas. Relatively smaller, but significant enhancement of the underlying systematic signal was achieved for the K and Ka samples.

\subsection{VSH fit of Gaia$-$S/X position offsets}
The iterative filtering and VSH fitting procedure generates a sample of 2146 S/X radio sources and their optical counterparts. The final 48 VSH coefficients and their covariances are collected in the $48\times 48$ symmetric matrix. We note that the fit may be equal to zero at a specific location on the sphere even when all of the fitting coefficients are nonzero. Table~1 lists all the 48 fitting coefficients $a_j$ in \uas\ along with the corresponding VSH identifications \{kind, phase, degree, order\}. The formal uncertainties $\sigma_j$ in \uas\ are given in column 4, and the S/N ratio $a_j/\sigma_j$ in column 5. The numbers given in Table~1 are the direct output of the LS fit for the original VSH functions, no re-scaling for the normalization VSH coefficients have been applied. We can see that the formal statistical significance of the fitting coefficients, as expressed via the S/N ratio, is rather small for most of the VSH terms. We find 17 terms in the fit with $|\text{S/N}|>2$, 12 above 2.5, 7 above 3.0, 1 above 5.0, and none above 10.0. Although the most significant terms appear among the lower degrees of the fit, signals with significant formal S/N ratios are found throughout the entire range of the fit. We note that a VSH fit is always approximate for a finite set of functions, because they are not orthogonal on the discrete set of data points. The results are to some degree dependent on the limiting order due to the internal dependencies of the fitted functions on the omitted functions \citep[i.e.,][]{2021A&A...649A...9G}. Within the set of 48 lower-degree harmonics, the largest correlation coefficient has an absolute value of 0.35. Alternatively, the ``leakage" between the VSH terms can be evaluated using the spectrum of singular values or the normalized robustness parameter, as discussed in Section \ref{pca.sec}. The most significant term is {\bf EVSH}$_{020}$, which is a zonal harmonic taking zero values at the celestial poles and the equator and vectors directed toward the equator at mid-latitudes.  The emergence of a first-degree magnetic term {\bf MVSH}$_{211}=\left\{\frac{1}{2} \sqrt{\frac{3}{2 \pi }} \sin (\alpha ) \sin (\delta ),\frac{1}{2} \sqrt{\frac{3}{2 \pi }} \cos (\alpha )\right\}$ among the first three VSH terms representing the rigid rotation of the Gaia frame with respect to the S/X frame may seem surprising, because the Gaia frame was rotated to coincide with the S/X part of ICRF3. We note, however, that considerable differences are acquired through a different, carefully vetted sample of reference sources, optimized weighting of condition equations, and using a larger set of fitting functions beyond the three rotation terms. The latter mitigates the harmonic leakage problem. The electric harmonics of first degree, on the other hand, are relatively insignificant. The Galactocentric acceleration of the Solar system barycenter should be present there, but it cannot be detected in this study. With a difference of only 1 year between the reference epochs of ICRF3 and Gaia DR3, the signal of the acceleration dipole is expected to be $\sim 5$ \uas\ in amplitude, which is much smaller that the other effects emerging from our analysis.

\begin{figure*}
    \includegraphics[width=0.78 \textwidth]{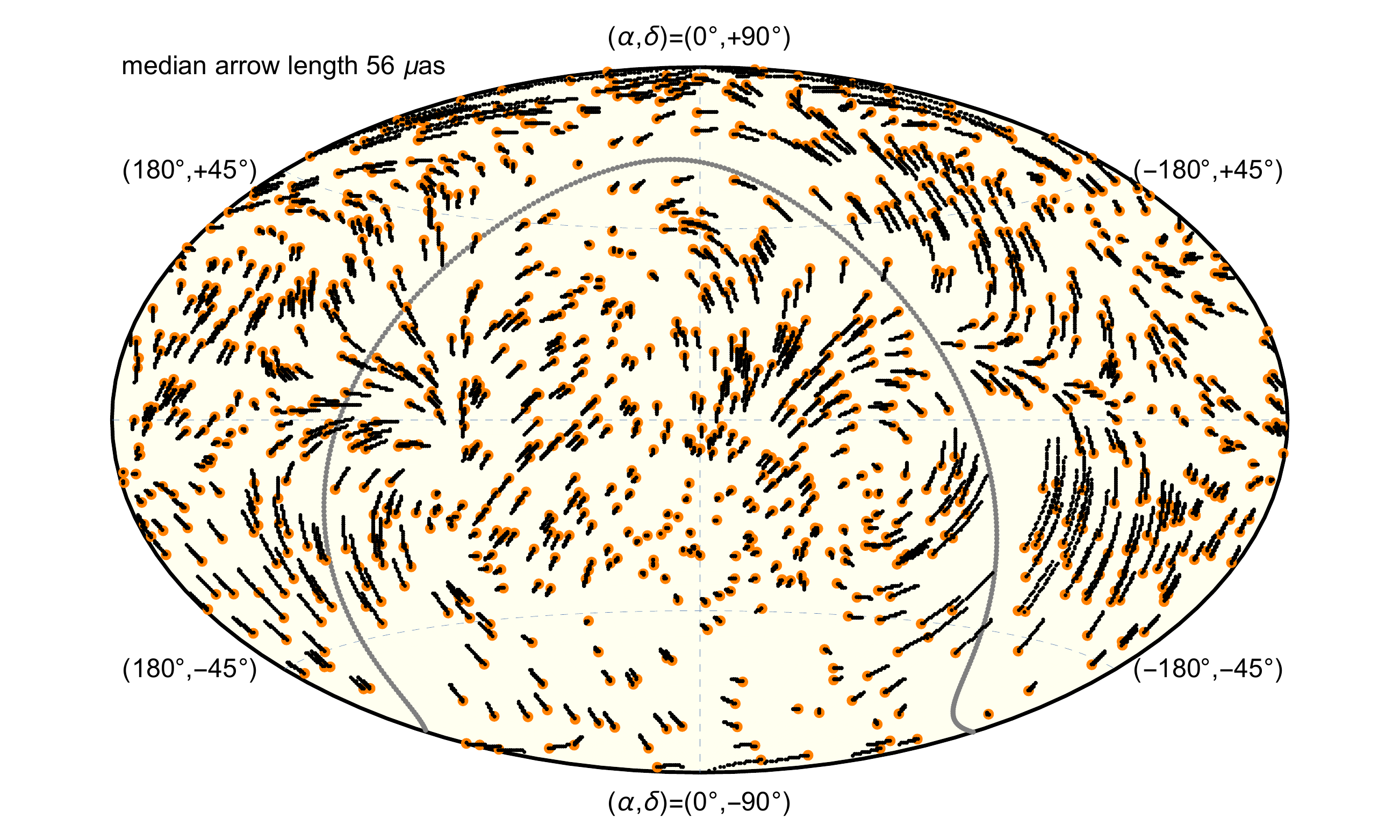}
    \caption{Graphical presentation of the VSH-fitted Gaia$-$S/X position offsets on the celestial sphere in the Aitoff equatorial projection. Orange dots at the origin of vectors indicate the actual position of S/X sources. The length and direction of small vectors represent the cumulative VSH fit for the corresponding point. The median length of the arrows is 56 \uas. The gray line shows the location of the nominal Galactic equator.}
    \label{sx.fig}
\end{figure*}

\begin{deluxetable*}{lc|rrr|rrr|rrr} \label{aj.tab}
\tabletypesize{\footnotesize}
\tablewidth{0pt}
\tablehead{
\colhead{} & \colhead{} & \multicolumn{3}{c}{S/X sample} &
\multicolumn{3}{c}{K sample} &\multicolumn{3}{c}{Ka sample} \\
\hline
\colhead{$j$} & \colhead{VSH id} & \colhead{$a_j$} & \colhead{$\sigma_j$} & \colhead{$a_j/\sigma_j$} & \colhead{$a_j$} & \colhead{$\sigma_j$} & \colhead{$a_j/\sigma_j$}& \colhead{$a_j$} & \colhead{$\sigma_j$} & \colhead{$a_j/\sigma_j$} \\
\colhead{} & \colhead{} & \colhead{$\mu$as} & \colhead{$\mu$as} &\colhead{} & \colhead{$\mu$as} &\colhead{$\mu$as} & \colhead{} &\colhead{$\mu$as} & \colhead{$\mu$as} & \colhead{}
}
\startdata
1 & \{\text{mag},0,1,0\} & 0.06 & 11.88 & 0.01 & -5.67 & 20.80 & -0.27 & 22.16 & 21.65 & 1.02 \\
 2 & \{\text{mag},1,1,1\} & 7.56 & 18.79 & 0.40 & 152.71 & 40.08 & 3.81 & -147.35 & 33.12 & -4.45 \\
 3 & \{\text{mag},2,1,1\} & -61.85 & 18.60 & -3.32 & -103.56 & 39.26 & -2.64 & -62.41 & 33.35 & -1.87 \\
 4 & \{\text{ele},0,1,0\} & -7.56 & 12.26 & -0.62 & 39.70 & 25.00 & 1.59 & -638.79 & 21.72 & -29.41 \\
 5 & \{\text{ele},1,1,1\} & -48.77 & 19.12 & -2.55 & -84.53 & 37.62 & -2.25 & -44.40 & 34.26 & -1.30 \\
 6 & \{\text{ele},2,1,1\} & -5.27 & 17.81 & -0.30 & 140.24 & 36.63 & 3.83 & 169.77 & 32.10 & 5.29 \\
 7 & \{\text{mag},0,2,0\} & -4.26 & 8.01 & -0.53 & 32.61 & 15.15 & 2.15 & -250.38 & 14.93 & -16.77 \\
 8 & \{\text{mag},1,2,1\} & 27.27 & 10.08 & 2.71 & 70.55 & 21.07 & 3.35 & 88.23 & 17.39 & 5.07 \\
 9 & \{\text{mag},2,2,1\} & -13.34 & 10.39 & -1.28 & 10.11 & 20.94 & 0.48 & -46.13 & 18.13 & -2.54 \\
 10 & \{\text{ele},0,2,0\} & -46.61 & 7.53 & -6.19 & -2.84 & 16.72 & -0.17 & -168.32 & 13.76 & -12.23 \\
 11 & \{\text{ele},1,2,1\} & -3.45 & 10.74 & -0.32 & -48.95 & 21.96 & -2.23 & -37.90 & 18.57 & -2.04 \\
 12 & \{\text{ele},2,2,1\} & -24.12 & 10.92 & -2.21 & 65.84 & 22.89 & 2.88 & -91.42 & 19.12 & -4.78 \\
 13 & \{\text{mag},1,2,2\} & 13.81 & 10.12 & 1.37 & -23.38 & 20.21 & -1.16 & 51.25 & 18.42 & 2.78 \\
 14 & \{\text{mag},2,2,2\} & 22.17 & 10.32 & 2.15 & 0.88 & 20.31 & 0.04 & -4.78 & 18.23 & -0.26 \\
 15 & \{\text{ele},1,2,2\} & -32.92 & 9.64 & -3.41 & 0.48 & 17.57 & 0.03 & -20.08 & 17.72 & -1.13 \\
 16 & \{\text{ele},2,2,2\} & 4.17 & 9.95 & 0.42 & -11.75 & 18.70 & -0.63 & -21.31 & 18.04 & -1.18 \\
 17 & \{\text{mag},0,3,0\} & 7.94 & 5.58 & 1.42 & -3.31 & 10.49 & -0.32 & -4.35 & 10.01 & -0.43 \\
 18 & \{\text{mag},1,3,1\} & -20.58 & 6.61 & -3.11 & 4.06 & 13.23 & 0.31 & -13.63 & 12.43 & -1.10 \\
 19 & \{\text{mag},2,3,1\} & 9.37 & 7.38 & 1.27 & 39.96 & 14.29 & 2.80 & 19.34 & 13.09 & 1.48 \\
 20 & \{\text{ele},0,3,0\} & -8.61 & 5.23 & -1.65 & 27.80 & 11.14 & 2.50 & -16.64 & 9.29 & -1.79 \\
 21 & \{\text{ele},1,3,1\} & -18.93 & 7.29 & -2.60 & -25.23 & 15.82 & -1.59 & 3.80 & 13.55 & 0.28 \\
 22 & \{\text{ele},2,3,1\} & -10.57 & 7.94 & -1.33 & 27.13 & 17.30 & 1.57 & -7.69 & 13.75 & -0.56 \\
 23 & \{\text{mag},1,3,2\} & -6.96 & 7.14 & -0.98 & -18.66 & 14.75 & -1.27 & -30.19 & 13.34 & -2.26 \\
 24 & \{\text{mag},2,3,2\} & 26.15 & 7.10 & 3.68 & 23.52 & 14.19 & 1.66 & 66.12 & 12.55 & 5.27 \\
 25 & \{\text{ele},1,3,2\} & 8.33 & 7.25 & 1.15 & 4.95 & 13.64 & 0.36 & 1.31 & 13.51 & 0.10 \\
 26 & \{\text{ele},2,3,2\} & 15.98 & 7.28 & 2.19 & -25.97 & 13.69 & -1.90 & -12.11 & 12.81 & -0.95 \\
 27 & \{\text{mag},1,3,3\} & -14.20 & 7.11 & -2.00 & -5.09 & 13.47 & -0.38 & 14.41 & 12.47 & 1.16 \\
 28 & \{\text{mag},2,3,3\} & 10.40 & 6.73 & 1.54 & 11.20 & 12.43 & 0.90 & 2.66 & 11.58 & 0.23 \\
 29 & \{\text{ele},1,3,3\} & 5.29 & 6.55 & 0.81 & -4.76 & 11.57 & -0.41 & 1.24 & 11.68 & 0.11 \\
 30 & \{\text{ele},2,3,3\} & 6.82 & 6.44 & 1.06 & -18.34 & 11.07 & -1.66 & -6.46 & 11.49 & -0.56 \\
 31 & \{\text{mag},0,4,0\} & -2.85 & 4.40 & -0.65 & -5.48 & 8.09 & -0.68 & -7.62 & 7.82 & -0.97 \\
 32 & \{\text{mag},1,4,1\} & -2.90 & 5.12 & -0.57 & -27.81 & 9.89 & -2.81 & -19.92 & 9.83 & -2.03 \\
 33 & \{\text{mag},2,4,1\} & 9.76 & 5.50 & 1.77 & -0.11 & 9.93 & -0.01 & 12.95 & 10.05 & 1.29 \\
 34 & \{\text{ele},0,4,0\} & 9.69 & 3.80 & 2.55 & 36.00 & 7.64 & 4.71 & 20.36 & 6.83 & 2.98 \\
 35 & \{\text{ele},1,4,1\} & -11.70 & 5.61 & -2.09 & -19.97 & 11.51 & -1.74 & 29.47 & 10.46 & 2.82 \\
 36 & \{\text{ele},2,4,1\} & -1.15 & 5.75 & -0.20 & -11.62 & 12.22 & -0.95 & 9.39 & 10.88 & 0.86 \\
 37 & \{\text{mag},1,4,2\} & -1.40 & 5.40 & -0.26 & -4.70 & 10.50 & -0.45 & 9.83 & 10.26 & 0.96 \\
 38 & \{\text{mag},2,4,2\} & -11.12 & 5.25 & -2.12 & -8.08 & 9.83 & -0.82 & -4.45 & 9.55 & -0.47 \\
 39 & \{\text{ele},1,4,2\} & -1.68 & 5.69 & -0.29 & 3.17 & 10.95 & 0.29 & 12.34 & 10.36 & 1.19 \\
 40 & \{\text{ele},2,4,2\} & -0.07 & 5.68 & -0.01 & -14.18 & 10.62 & -1.33 & -19.88 & 10.18 & -1.95 \\
 41 & \{\text{mag},1,4,3\} & -4.51 & 5.28 & -0.85 & -20.51 & 10.14 & -2.02 & 11.80 & 9.84 & 1.20 \\
 42 & \{\text{mag},2,4,3\} & 20.23 & 5.37 & 3.77 & 3.06 & 10.25 & 0.30 & 3.44 & 9.61 & 0.36 \\
 43 & \{\text{ele},1,4,3\} & -7.01 & 5.12 & -1.37 & 4.36 & 9.18 & 0.47 & -1.69 & 9.55 & -0.18 \\
 44 & \{\text{ele},2,4,3\} & -2.30 & 5.44 & -0.42 & 2.97 & 9.41 & 0.32 & -5.98 & 9.72 & -0.61 \\
 45 & \{\text{mag},1,4,4\} & 7.21 & 5.15 & 1.40 & 10.51 & 9.38 & 1.12 & -6.82 & 8.71 & -0.78 \\
 46 & \{\text{mag},2,4,4\} & -7.00 & 5.44 & -1.29 & 2.42 & 9.56 & 0.25 & 8.71 & 9.09 & 0.96 \\
 47 & \{\text{ele},1,4,4\} & 14.55 & 4.74 & 3.07 & 18.15 & 7.84 & 2.31 & -4.42 & 8.13 & -0.54 \\
 48 & \{\text{ele},2,4,4\} & -13.67 & 4.96 & -2.76 & -23.20 & 8.01 & -2.90 & -7.01 & 8.45 & -0.83 \\
\enddata
\caption{Numerical results of the VSH fitting for the S/X, K, and Ka samples of Gaia$-$ICRF3 position offsets.}
\end{deluxetable*}

Figure~\ref{sx.fig} shows the final S/X VSH fit in graphical form. The small vectors show the length and direction of the fit (Equation~\ref{fit.eq}) for each S/X source at its specific position indicated by an orange dot. The common scale of the offset vectors is given for the median length in the upper left corner. The nominal vernal equinox (origin of the equatorial coordinate system) is at the center of the plot, and R.A.\ increases from right to left. The manifestation of the dominating harmonic number 10 is the prevalence of downward arrows in the northern hemisphere and upward arrows in the south. However, the combined pattern of the fit is quite non-uniform and complex. The largest signal is confined to the area in the third quadrant in R.A.\ between decl.~$-45\arcdeg$ and the equator. This local streaming pattern is well aligned with the adjacent Galactic plane. The complexity of the field is related to the fact that relatively large terms appear across the entire range of the fit (with the second most significant term {\bf MVSH}$_{243}$ being of degree 4). The central area of the plot around the vernal equinox, on the other hand, is characterized by relatively small systematic offsets. 

\begin{figure*}
    \includegraphics[width=0.78 \textwidth]{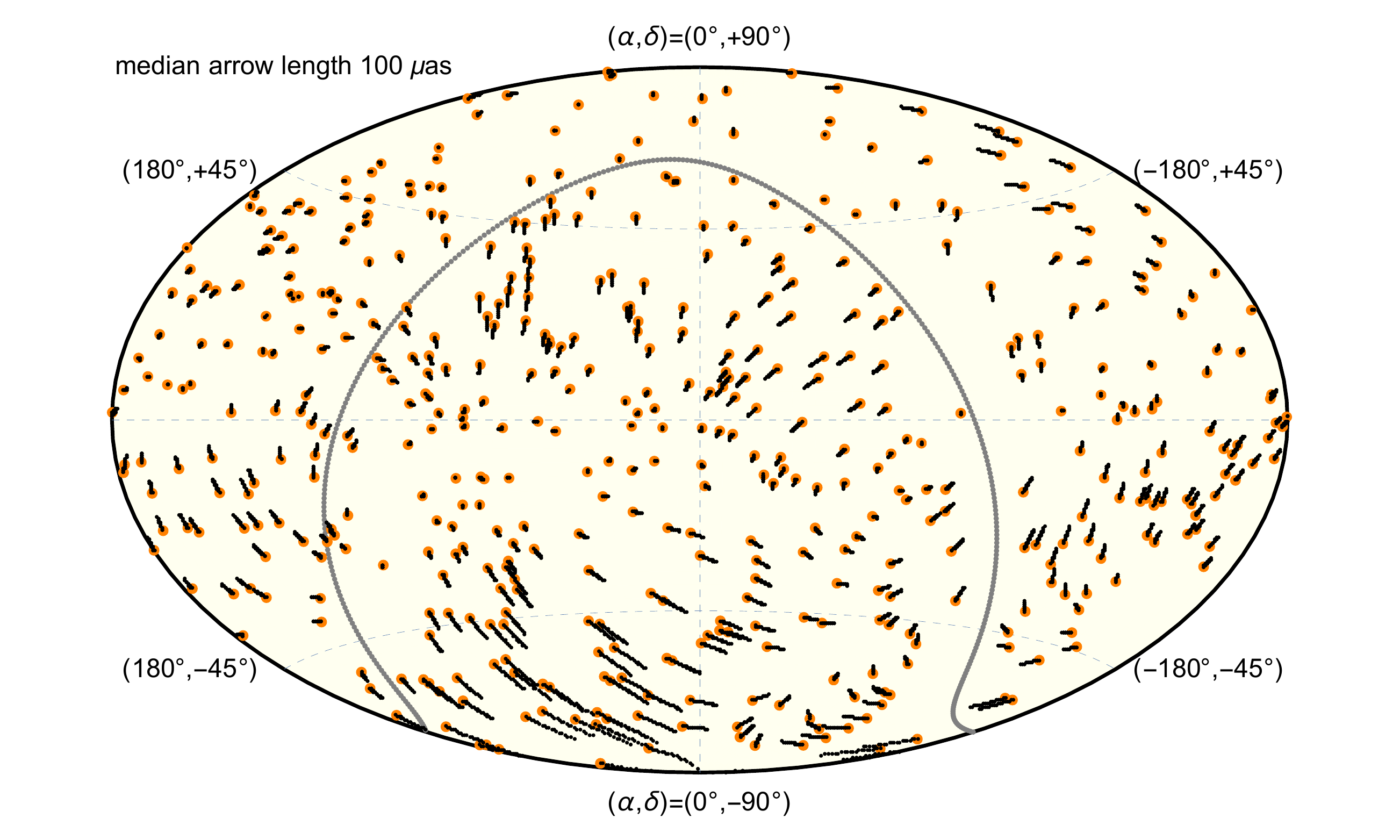}
    \caption{Graphical presentation of the VSH-fitted Gaia$-$K position offsets on the celestial sphere in the Aitoff equatorial projection. Orange dots at the origin of vectors indicate the actual position of K sources. The length and direction of small vectors represent the cumulative VSH fit for the corresponding point. The median length of the arrows is 100 \uas. The gray line shows the location of the nominal Galactic equator.}
    \label{kx.fig}
\end{figure*}

\subsection{VSH fit of Gaia$-$K position offsets}
We used exactly the same procedure and three full-scale iterations to obtain a VSH fit of the Gaia$-$K position offsets. The results are given in Table~1 as the fitting coefficients $a_j$, their formal errors 
$\sigma_j$, and the S/N ratio $a_j/\sigma_j$, in columns 6--8. The formal uncertainties of the $a_j$ coefficients for the K fit are roughly twice as large as those for the S/X fit. This is the direct consequence of the smaller number of sources in the K sample. The conditioning of the design matrix, which is related to the eigenvector spectrum (Sect. \ref{pca.sec}), also played a minor role in this relative underperformance. Despite the more uncertain estimates, the number of statistically 
significant terms is slightly smaller that of the Gaia$-$S/X field, with 15 terms in the fit with $|\text{S/N}|>2$, 9 above 2.5, 4 above 3.0, and none above 5.0. Overall, the K sample shows stronger sky-correlated perturbations than the S/X sample. We note the emergence of a marginally significant term {\bf MVSH}$_{111}=\left\{\frac{1}{2} \sqrt{\frac{3}{2 \pi }} \cos (\alpha ) \sin (\delta ),-\frac{1}{2} \sqrt{\frac{3}{2 \pi }} \sin(\alpha )\right\}$, which represents a negative right-handed rotation of the Gaia frame with respect to the K frame around the first axis toward the vernal equinox. Importantly, the strongest detected signal {\bf EVSH}$_{020}$ for the S/X field is not present at all in the K field. Instead, a first-order dipole term {\bf EVSH}$_{211}$ emerges, which was quite insignificant in the S/X field. The statistically dominant term is the zonal harmonic {\bf EVSH}$_{040}$, which is less prominent in the S/X field.

Figure~\ref{kx.fig} shows the distribution of the fitted Gaia$-$K offset field in the sky projection. There is still a hint of systematic upward pattern in the southern third quadrant of the sky around the Galactic plane, but it is dwarfed by the most prominent feature, which the streaming pattern toward the south pole in the southern cap, first and second quadrants. This pattern is counter to the field fitted for the S/X sample (Figure~\ref{sx.fig}). The similarities between the two fits are limited to the central area around the origin of the equatorial system, where the terms combine to produce a nearly zero offset. Generally, the strongest signal in the K field is more localized on the sky than in the S/X field. The non-repeatability of the offset patterns between the different radio wavelength bands is the main result of this analysis. We also note the prevalence of features correlated with the celestial equator (rather than the Galactic plane) in both samples, which points at a technical origin of the prevalent pattern in the VLBI positions.

\subsection{VSH fit of Gaia$-$Ka position offsets}
\begin{figure*}
    \includegraphics[width=0.78 \textwidth]{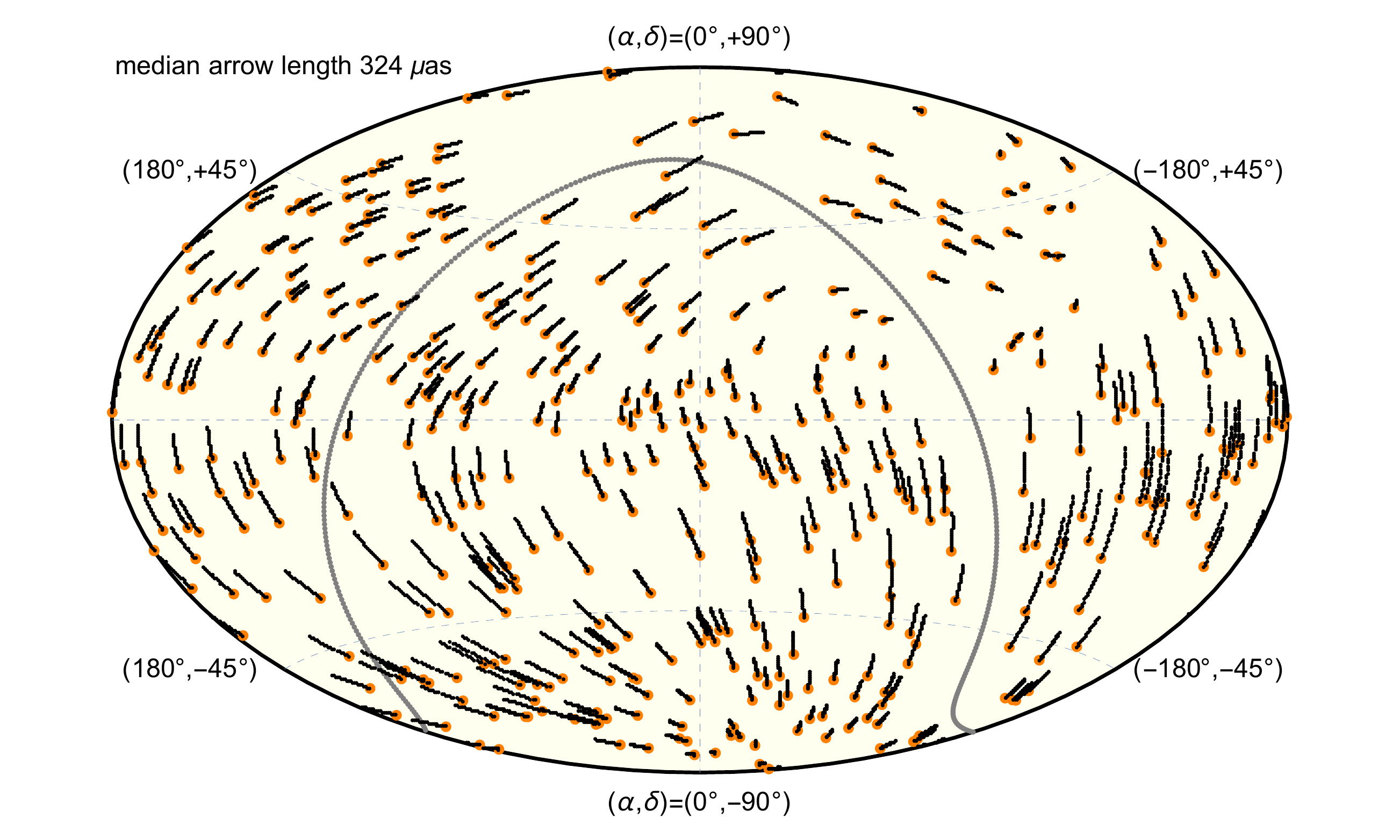}
    \caption{Graphical presentation of the VSH-fitted Gaia$-$Ka position offsets on the celestial sphere in the Aitoff equatorial projection. Orange dots at the origin of vectors indicate the actual position of Ka sources. The length and direction of small vectors represent the cumulative VSH fit for the corresponding point. The median length of the arrows is 324 \uas. The gray line shows the location of the nominal Galactic equator.}
    \label{ka.fig}
\end{figure*}

An independent fit of the Gaia$-$Ka position offsets was performed on the 432 sources of the selected precious set using the same algorithm and the limiting VSH degree. The iterative selection required five iterations in this case, which increased the sample by 125 sources compared to the initial cut. This already indicates the presence of large systematic offsets in the data. The results are presented in Table~1, columns 9--11. The formal uncertainties of the VSH coefficients are roughly 3 times the errors of the S/X sample. This is explained by the smaller number of sources and a weaker conditioning related to the distribution on the sky. The formal single measurement errors, on the other hand, are not drastically larger for the Ka sample. The fitting coefficients are remarkably larger both in absolute and relative value, and we obtain more highly significant terms. Fifteen terms in the fit are found to have $|\text{S/N}|>2$, 12 above 2.5, 8 above 3.0, 6 above 5.0, and 3 above 10.0. The dominating VSH term is {\bf EVSH}$_{010}=\left\{0,-\frac{1}{2} \sqrt{\frac{3}{\pi }} \cos (\delta )\right\}$, which is a simple dipole streaming from north to south. It is so much larger than the rest of the terms that its pattern is obvious even in the combined fit in Figure~\ref{ka.fig}. The other highly significant terms are {\bf MVSH}$_{020}=\left\{\frac{3}{2} \sqrt{\frac{5}{\pi }} \sin (\delta ) \cos (\delta ),0\right\}$ (antisymmetric rotation around the poles) and {\bf EVSH}$_{020}=\left\{0,\frac{3}{2} \sqrt{\frac{5}{\pi }} \sin (\delta ) \cos (\delta )\right\}$ (antisymmetric glide toward the equator). The latter term is fairly consistent with the greatest pattern in the S/X field, but the similarities between the samples do not extend further.  The median length of the fitted offset vector is 324 \uas. It is much larger than the median formal error of K positions in the precious set (111 \uas\ per coordinate). Therefore, the accuracy of this part of ICRF3 can be drasticaly improved by subtracting the specific sky-correlated distortion found in this paper.

\section{Robustness of VSH fits and principal component analysis}
\label{pca.sec}
How reliable are the presented VSH fits for the three constituents of ICRF3 matched with the optical CRF?
A quantitative answer is obtained by the singular value decomposition of the normal matrix and computation of the least-squares (LS) robustness parameter $\cal{R}$ introduced in \citep[][Equation~13]{2021AJ....161..289M}. While the robustness of an ideal, perfectly conditioned problem is defined by the ratio of the number of equations to the number of unknowns, real LS problems are additionally weakened by the propagation of error along the least significant eigenvectors. The spectrum of singular values thus defines how well-conditioned the solution is with respect to random and arbitrary systematic errors in the right-hand part of the equations. A well-conditioned solution has a flat spectrum, i.e., a slowly declining sequence of singular values $s_j$, $j=1,2,\ldots,m$. To separate the structural strength from the robustness related to the number of condition equations $n$ with respect to the number of unknowns $m$, we introduce a {\it normalized robustness} parameter,
\eb
\bar{\cal R}\equiv {\cal R}\,\sqrt{m}/\sqrt{n}=\frac{\sqrt{m}}{\sqrt{\sum_{j=1}^m s_1^2/s_j^2}}.
\ee
This positive definite number is less than or equal to 1, and it can be interpreted as the structural weakness of a given LS solution caused by the steepness of the singular value spectrum. A value $\bar{\cal R}$ much smaller than 1 signifies a structurally weak solution. 

\begin{figure*}
    \includegraphics[width=0.78 \textwidth]{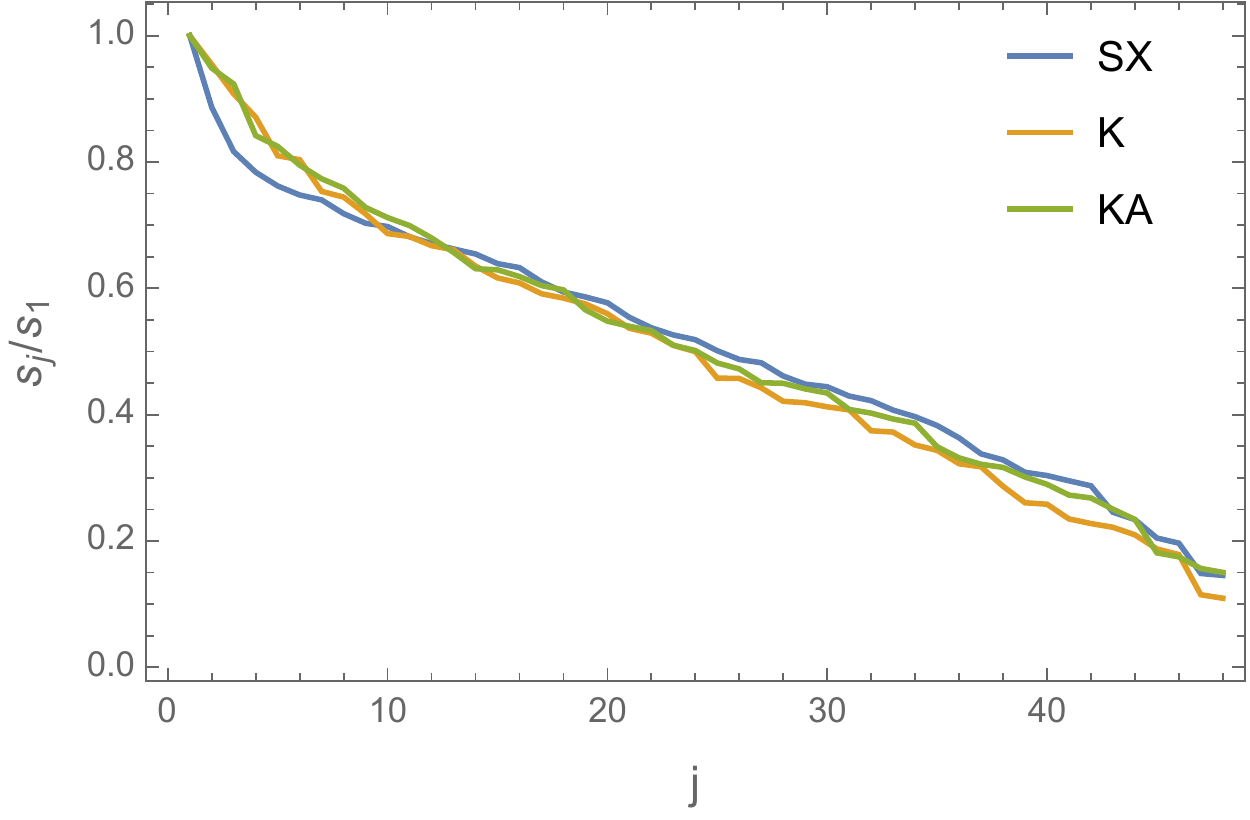}
    \caption{Singular values of the VSH fits for the three radio-optical cleaned samples (S/X, K, Ka) normalized to the greatest singular value $s_1$.}
    \label{sj.fig}
\end{figure*}

Figure~\ref{sj.fig} shows the 48 singular values normalized by the greatest singular value, $s_j/s_1$, for each of the three sets of radio-optical reference frame objects. The lower the curve, the weaker (i.e., less robust) is the solution. If $\bar{\cal R}$ drops to low values, the setting should be changed to improve the robustness. In our case, this can in principle be achieved either by reducing the limiting degree $L$ of the VSH fit or by boosting the number of sources and improving the sky coverage. The singular value spectra of the three samples are fairly close despite the differences in the numerical fits. This consistency emerged in the iterative selection/fitting process, as the initial iteration produced a much greater spread of singular values. Consequently, the normalized robustness increased significantly for all three parts, especially for the Ka sample. The directly computed values are $\bar{\cal R}_{\rm S/X}=0.36$,
$\bar{\cal R}_{\rm K}=0.31$, and $\bar{\cal R}_{\rm Ka}=0.36$. The somewhat lower performance of the K fit is probably related to a less favorable configuration of selected reference sources in the southern cap. Using a higher-degree decomposition is hardly possible because it would weaken the VSH solutions.

\section{Considerations on the catalog combination}\label{disc.sec}
The vector fields of position offsets Gaia$-$ICRF3 generally include two components, which can be called accidental and systematic errors. For the radio-optical position offsets considered in this paper, the accidental part comes from the stochastic, uncorrelated error of the measurement (ultimately, from the underlying Poisson noise of the registered photon counts and the instrument/detector noise), and from the astrophysical---i.e., genuine---wavelength-dependent displacements of photocenters. The latter are assumed to be uncorrelated between the sources, although this cannot be experimentally proven. Therefore, they are stochastic only on a sufficiently large sample of sources, being fully correlated in multiple observations of a single source. The systematic  (sky-correlated) part, which is shared by nearby sources, is assumed to be completely man-made, originating in deficiencies of the data reduction algorithms and inadequacies of adopted data models. The systematic error has a great variety of possible dependencies (e.g., source brightness, morphology, color, local configuration of the baselines or scan directions), which may be impossible to completely identify. The dissimilarity of the most significant VSH terms between the three ICRF3 components indicates that each of these components has its own wavelength-specific systematic distortion, despite the previous indications of the opposite \citep{2019A&A...630A.101K}.

Our aim is to construct a combined reference frame that is more precise and accurate than each of the components. This possibility and different approaches have been discussed, e.g., by \citet{2007A&A...474..665S}. The random part of the error can be diminished by simple weighted averaging of the available positions for each source. If the weights were the same and a source is present in all the four catalogs, the formal uncertainty, which nominally captures the accidental error budget, would be reduced by 50\%. A source only found in two of the catalogs would obtain a more modest boost of the expected precision of up to 29\%. This procedure, however, does not improve the result in the systematic sense. Furthermore, it can muddle the resulting frame because, depending on the relative weights of the four position vectors, the systematic fields will propagate differently among the sources. The simplest solution would be to bring the systematic fields to a single system, for example, to that of the ICRF3 S/X catalog. This choice would also be justified by the fact that the Gaia DR3 catalog has been already adjusted to that system as far as the three VSH components responsible for rigid rotation are concerned. 

In this study, we explore a different approach. Undoubtedly, the S/X radio catalog has its own set of systematic errors, which is not guaranteed to be smaller than the other three components. We posit that additional improvement can be expected in a slightly more elaborate approach based on the comparison of the computed VSH terms. Consider a marginal case when the three ICRF3 catalogs have negligible systematics and the Gaia catalog has some detectable distortions. Then, within the random errors of the VSH determination (described by the formal errors in Table~1), the numerical values of the VSH coefficients would be the same between the three differential solutions. In an opposite marginal case, the Gaia catalog is free of systematic errors, then the VSH coefficients represent the radio band-specific systematic fields. In reality, we have a mixture of Gaia-specific and ICRF3-specific systematics in the results in Table~1. Only 14 VSH terms out of the total 48 fitted terms are significant ($|$S/N$|>3$) in at least one sample. Of those, only 4 terms are significant in two samples, and none are significant in all three samples. Only 3 VSH terms that are significant in two samples actually have the same sign (namely, number 8, 10, and 24 in Table 1). This strongly suggests that the systematic vector fields are essentially disparate between the three samples, even though the radio catalogs were observed with mostly the same instrument (VLBI). It is therefore reasonable to propose that the ``average" part of the three VSH coefficients for each statistically insignificant term is closer to the Gaia-specific error, whereas a significant deviation from it is a feature of the corresponding ICRF3 catalog. The statistically insignificant terms may still be partially correlated between the three samples because of a common systematic field in the ICRF3 catalogs, which is below the detectable level. The intrinsic ambiguity of its origin cannot be resolved with the available data. Subtracting the mean VSH field from the Gaia positions for the statistically insignificant terms is logically consistent with the adopted strategy of Gaia CRF construction, where the Gaia frame is aligned with the ICRF3 frame within the three VSH terms of rigid rotation. Our adopted approach is an extension of this strategy onto the larger set of VSH functions.
Thus, separating the mean and the deviation of the estimated values for each term and treating statistically significant and discordant terms separately should allow us to improve the system of the combined reference frame.

\section{Construction of the Radio-Optical Reference Catalog}\label{con.sec}
The combined RORC-1 includes the union of the three precious sets S/X, K, and Ka cross-matched with the Gaia DR3 catalog. The construction procedure is divided into two steps:
\begin{enumerate}
\item
Correct the systematic fields of all four catalogs using the VSH decomposition results in Table~1.
\item
Compute a weighted average position for each source using all available corrected catalog positions and their formal covariances. This average position is the main derived data in RORC-1.
\end{enumerate}
The first step begins with computing the average Gaia$-$ICRF3 field by
\eb
\bar a_j=\sum a_{jq}\, \sigma^{-2}_{jq}\,/ \sum \sigma^{-2}_{jq}.
\ee
The summation over $q=$\{S/X,K,Ka\} picks only those coefficients, which have absolute S/N ratios smaller than 3. The omitted values are believed to represent significant radio band-specific systematics, which should not be involved in the calculation of the mean field. All three values are involved in this computation for most of the terms. Ten terms are averaged on two (e.g., S/X and K) values only eliminating the statistically significant results for the third band, and in four cases, including the {\bf EVSH}$_{020}$ term discussed above, the mean is equal to the value determined for one of the samples. 

The mean field $\bar{\boldsymbol{d}}(\alpha,\delta)$ is computed using the mean VSH coefficients by Equation~\ref{fit.eq}. This mean systematic field is {\it subtracted} from the Gaia positions for each RORC-1 source. We then compute the deviations $a_{jq}-\bar a_j$ for each of the three samples S/X, K, and Ka. They provide three differential vector fields $\boldsymbol{d}_q$, which are {\it added} to the positions of sources in the corresponding parts of ICRF3. The VSH fits are thus fully utilized bringing the frames of Gaia and the three parts of ICRF3 to a common system.

The second step of this algorithm deals with individual corrected position vectors $\boldsymbol{r}_q$ for each source from the union of samples Gaia, S/X, K, and Ka. The smallest number of positions for a given source is 2 (Gaia and one of the ICRF3 parts), and the largest is 4. Let $\boldsymbol{r}_q$ be the 3D position vectors of unit length for a given source, with $q$ picking the available catalogs in the sequence \{Gaia, S/X, K, Ka\}. The average position vector is derived from the Maximum Likelihood principle, which implicitly assumes that the input catalog positions are independent and binormally distributed:
\begin{eqnarray}
\boldsymbol{r}_0 &=&\boldsymbol{r}_{\rm ref} + \boldsymbol{\rho}\\ \nonumber
\boldsymbol{\rho} &=&
\boldsymbol{C}_0\;\sum_q \boldsymbol C_q^{-1} \boldsymbol{\rho}_q,
\end{eqnarray}
where the tangential offsets $\boldsymbol\rho_q=\{(\boldsymbol{r}_q-\boldsymbol{r}_{\rm ref})\cdot \boldsymbol{\tau}_x, (\boldsymbol{r}_q-\boldsymbol{r}_{\rm ref})\cdot \boldsymbol{\tau}_y\}$, with $\boldsymbol{\tau}_x$ and $\boldsymbol{\tau}_y$ being the east- and north-directed coordinate vectors in the local tangent plane. This computation is invariant to the choice of the reference position $\boldsymbol{r}_{\rm ref}$, as long as it is close to the input catalog points.
$\boldsymbol C_0$ is the formal covariance matrix of the average position:
\eb
\boldsymbol{C}_0=\left(\sum_q \boldsymbol{C}_q^{-1}\right)^{-1}.
\ee

\begin{figure*}
    \includegraphics[width=0.48 \textwidth]{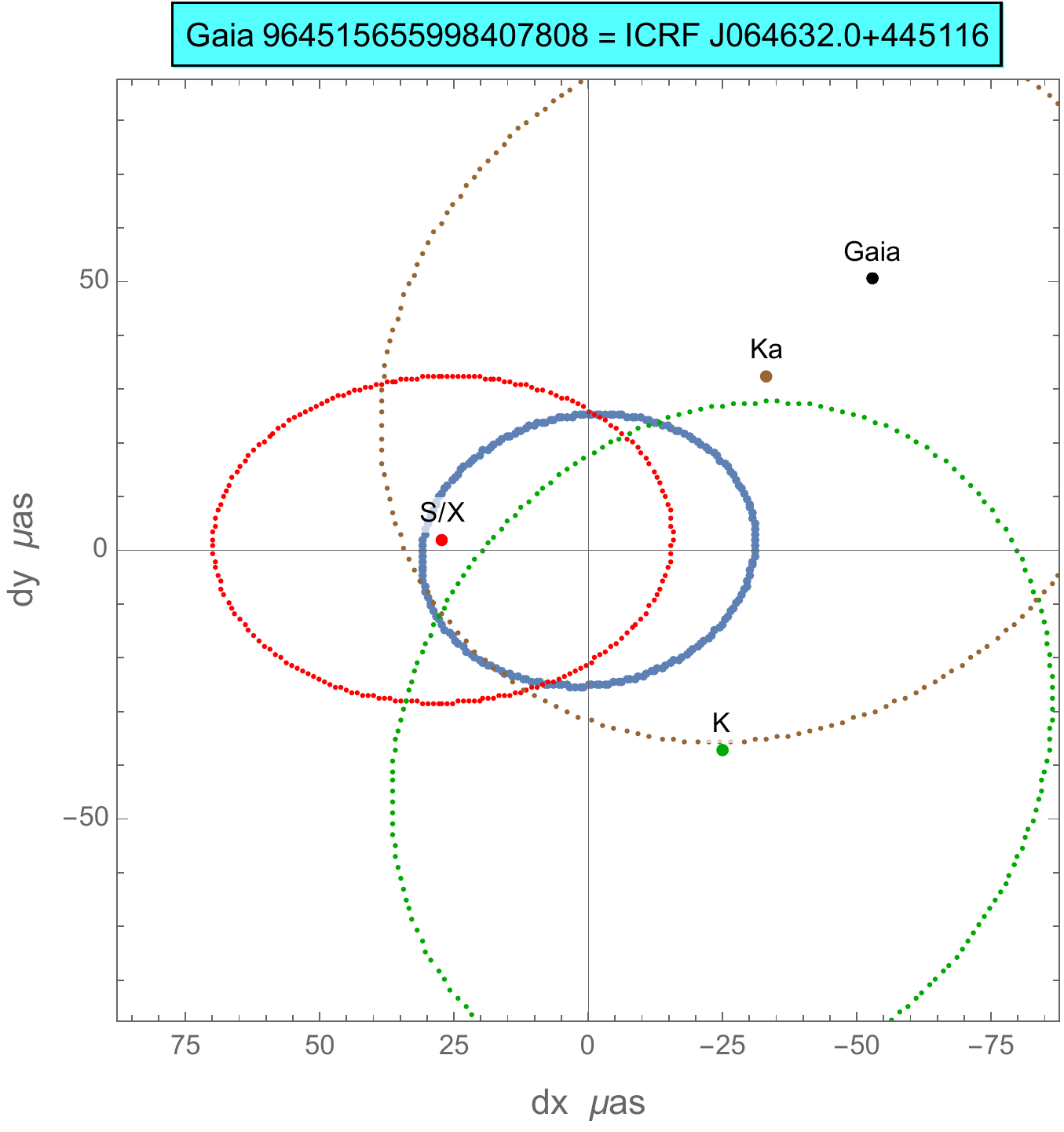}
    \includegraphics[width=0.48 \textwidth]{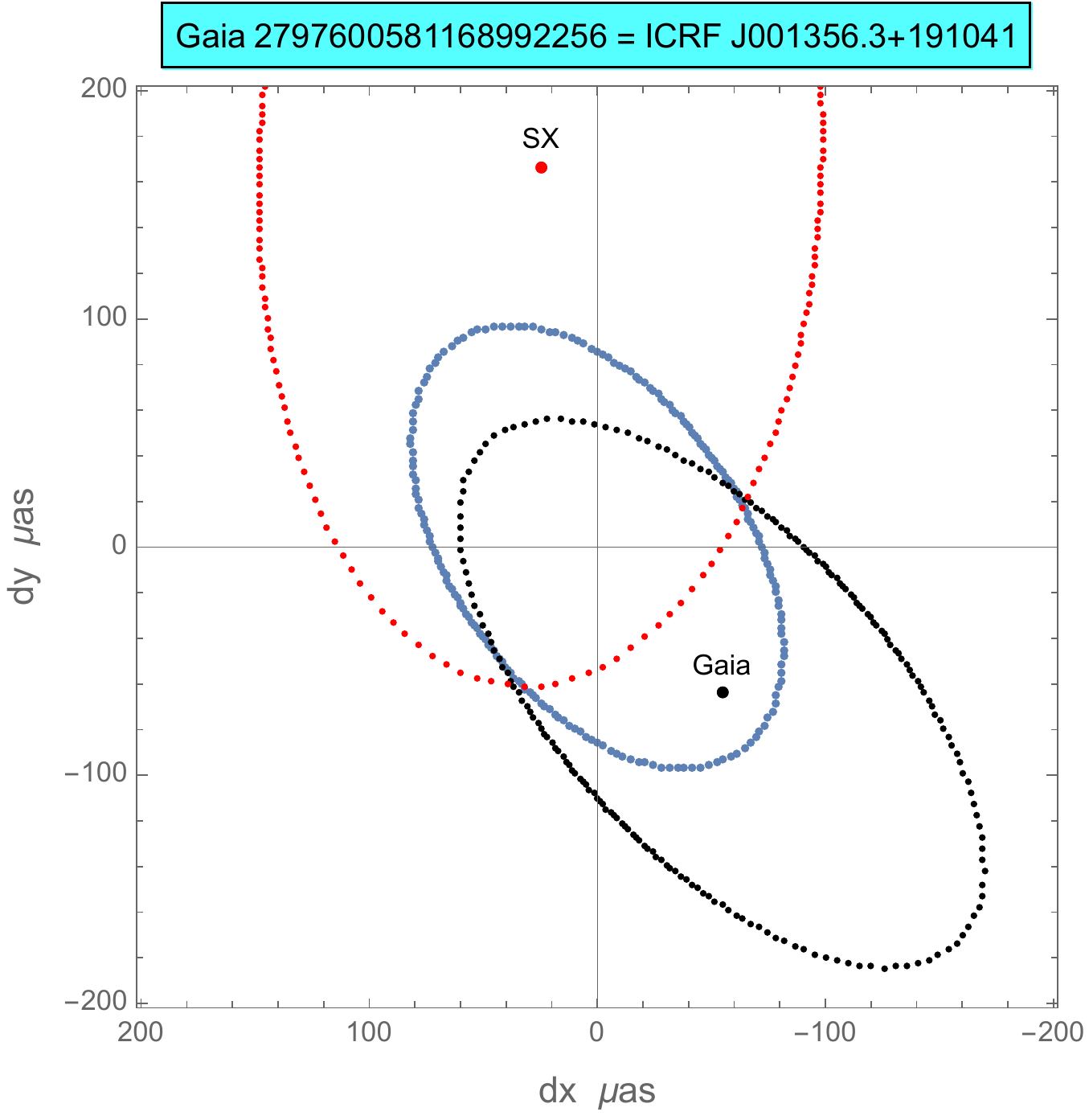}
    \caption{Configuration plots for two sources showing the derived RORC-1 position at the center and its error ellipse (blue thick line) and the VSH-corrected positions with their error ellipses from the available input catalogs Gaia, S/X, K, and Ka, as indicated in the legend.}
    \label{disp.fig}
\end{figure*}

Following this algorithm, we generated a combined RORC-1 catalog, which includes 2284 sources. This number is larger that the size of the S/X sample by only 138, because most of the sources in the K and Ka samples are the same as in S/X. Indeed, we find 299 sources that are present in all the three samples, 397 sources in common between S/X and K, and 337 sources in common between S/X and Ka. 38 sources are present
only in the Ka sample. Figure~\ref{disp.fig} illustrates two configurations of input (VSH-corrected) multi-frequency positions, the derived RORC-1 position, and their formal covariances on the sky. The tangent plane coordinates $x$ and $y$ are in the east (increasing R.A.) and north (increasing decl.) directions. The source Gaia 964515655998407808 in the left plot is one of the 299 RORC-1 entries with all four input positions available. The average RORC-1 position is at the origin of the tangent plane, and its error ellipse is shown as a blue dotted ellipse. The semiaxes of this error ellipse are 25 and 31 \uas. The four input positions are shown with labeled dots at the centers of the corresponding formal error ellipses. We can see that for this source, the uncertainty of the Gaia position is much larger than the three radio positions, which mostly define the result. The averaged position is offset by less than 1 standard deviation from the most precise S/X position, which obtains the highest weight. It is also within the error ellipses of the K and Ka positions. The other example in the right panel for Gaia 2797600581168992256 represents a case when only two input positions are available. The result is closer to the formally more precise Gaia position (black dot and ellipse). We note that the formal errors of Gaia astrometry are strongly dependent on the optical magnitude of the source, and in many cases, they are comparable to, or even smaller, than the ICRF3 formal errors. The semimajor axes of the RORC-1 position are 64 and 109 \uas.

\begin{deluxetable*}{lcclr} \label{format.tab}
\tablehead{\colhead{column} & \colhead{parameter} & \colhead{unit} & \colhead{meaning} & \colhead{note} }
\startdata
1 & Gaia source id &  &   & from Gaia DR3 \\
2 & ICRF name      &  &   & from ICRF3 catalogs\\
3 & R.A.\ & degrees & right ascension  & derived RORC-1 position\\
4 & decl.\ & degrees & declination  & derived RORC-1 position\\
5 & $\sigma_{\alpha *}$  & \uas\ & formal error in R.A.\ &  \\
6 & $\sigma_{\delta}$ & \uas\ & formal error in decl.\ &  \\
7 & $\rho$ &     & R.A.-decl.\ correlation &  \\
8 & \texttt{f\_G} &  & availability in Gaia DR3 & always set to 1 \\
9 & \texttt{f\_S/X} &  & availability in S/X & 0 if not available, 1 if available\\
10 & \texttt{f\_K} &  & availability in K & 0 if not available, 1 if available\\
11 & \texttt{f\_Ka} &  & availability in Ka & 0 if not available, 1 if available
\enddata
\caption{Format  of RORC-1 catalog (published online).}
\end{deluxetable*}

Table~\ref{format.tab} summarizes the format and contents of RORC-1. The catalog includes 11 columns and 2284 rows. The first two columns are the Gaia source identification number and the matching ICRF3 name. The derived average and corrected equatorial coordinates in R.A.\ and decl.\ are provided in columns 3 and 4, respectively, and their formal errors in columns 5 and 6. The correlation coefficient of these derived coordinates is given in column 7. The remaining four columns are binary flags (0 or 1) that specify the availability of input positions in this order of input samples: Gaia, S/X, K, and Ka. Since a Gaia position is always involved in the computation by construction, field \texttt{f\_G} in column 8 is always set to 1. There are no empty fields in the catalog. Table \ref{rorc.tab} gives the first ten rows of RORC-1. The entire catalog is published online. Note that the output coordinates are specified in degrees rounded to 10 decimal places, while their formal errors are given in integer \uas.

\begin{deluxetable*}{lcccccccccr} \label{rorc.tab}
\tablehead{\colhead{ICRF name} & \colhead{Gaia source id} & \colhead{R.A.} & \colhead{decl.} & \colhead{$\sigma_{\alpha *}$} &
 \colhead{$\sigma_{\delta}$} & \colhead{$\rho$} &\colhead{\texttt{f\_G}} & \colhead{\texttt{f\_S/X}} &\colhead{\texttt{f\_K}} & \colhead{\texttt{f\_Ka}}}
\startdata
  \text{ICRF J030123.6+011835} & 239556096283136 & 45.3483624387 & 1.3099989516 & 92 & 151 & -0.117 & 1 & 1 & 0 & 0 \\
 \text{ICRF J025850.5+054108} & 6884282820204288 & 44.7105264111 & 5.6855670374 & 93 & 166 & -0.172 & 1 & 1 & 0 & 0 \\
 \text{ICRF J030133.7+060227} & 6993551083342976 & 45.3904745662 & 6.0409117496 & 93 & 134 & 0.013 & 1 & 1 & 0 & 0 \\
 \text{ICRF J030903.6+102916} & 15495005053467776 & 47.2650978884 & 10.4878724802 & 26 & 30 & -0.154 & 1 & 1 & 1 & 1 \\
 \text{ICRF J032153.1+122113} & 17475775250904960 & 50.4712646323 & 12.3538760346 & 39 & 56 & -0.146 & 1 & 1 & 0 & 0 \\
 \text{ICRF J023714.0+052649} & 18121326015612672 & 39.3084918028 & 5.4472138082 & 167 & 193 & -0.2 & 1 & 1 & 0 & 0 \\
 \text{ICRF J023204.8+072605} & 19570173103308416 & 38.0200064750 & 7.4348594420 & 105 & 106 & 0.137 & 1 & 1 & 0 & 0 \\
 \text{ICRF J022053.8+094535} & 24069649561839104 & 35.2241762775 & 9.7597717596 & 188 & 177 & -0.005 & 1 & 1 & 0 & 0 \\
 \text{ICRF J022541.9+113425} & 24740523453770368 & 36.4246245082 & 11.5737402525 & 107 & 110 & 0.003 & 1 & 1 & 0 & 0 \\
 \text{ICRF J024229.1+110100} & 25350786766710528 & 40.6215451999 & 11.0168688664 & 29 & 42 & -0.289 & 1 & 1 & 1 & 1 \\
\enddata
\caption{First 10 lines of RORC-1 catalog (published entirely online). For detailed description of columns contents and format see Table \ref{format.tab}.}
\end{deluxetable*}

To illustrate the gain in precision achieved in RORC-1 relative to the input catalogs, we calculate the quantiles of the coordinate formal errors. The 0.5 quantiles (medians) for R.A.\ and decl.\ formal errors are, respectively, 181 and 167 \uas\ for Gaia, 143 and 202 \uas\ for S/X, 95 and 148 \uas\ for K, and 111 and 111 \uas\ for Ka. The 0.5 quantiles of the coordinate errors in RORC-1 are 90 and 98 \uas. Although the median formal errors are comparable in RORC-1 and the smaller K and Ka samples, significant improvements are seen for Gaia and the most important part of ICRF3 (S/X). An even greater advance in formal precision is found for the 0.9 quantiles of coordinate formal errors: 610 and 535 \uas\ for Gaia, 651 and 624 \uas\ for S/X, 563 and 460 \uas\ for K, 583 and 323 \uas\ for Ka, and 181 and 222 \uas\ for RORC-1. Thus, the number of sources with relatively large uncertainties is drastically reduced in RORC-1 compared to the input samples. 

\section{Conclusions}
\label{end.sec}
We have shown that the optical Gaia CRF can be combined with the three parts of the radio CRF realized through ICRF3 resulting a multi-frequency CRF that is likely to surpass the input constituents both in systematic accuracy and accidental precision. The main motivation for this work is the possibility to separate, in the statistical sense, the VLBI-specific systematic and sky-correlated errors from the distortions of the Gaia CRF, which result from random perturbations and calibration issues of this space mission. The main risk is associated with the now proven occurrence of genuine, i.e., physical, differences between the radio and optical positions of AGNs and quasars. With about one-quarter of the largest ICRF3 catalog (S/X) being affected by these offsets, the emerging problem is how to select most stable astrometric reference sources that are either free of these adverse effects, or have physical offsets below the currently achieved level of accuracy. The detectable astrometric offsets are manifestations of a multitude of astrophysical and instrumental effects, such as the core shift in active radio jets, the asymmetric and extended structures surrounding nearby AGNs, dual quasars and chance alignments with foreground stars, gravitational lensing, etc. As our knowledge of contributing factors grows, educated and targeted selection prescriptions can be offered. However, these prescriptions turn out to be either overly restrictive, or requiring additional astrophysical information that is currently unavailable. For example, highly variable blazar sources have recently been shown to exhibit optical-radio position offsets far less frequently than their less variable radio-loud counterparts \citep{2022ApJ...939L..32S}, probably due to the smaller line-of-sight angles of their radio jets. Unfortunately, the fraction of blazars in ICRF3 (and generally on the sky) is too small to be exclusively used as a basis of a multi-frequency CRF. For this first version of RORC, we employ a heuristic approach selecting radio-optical sources mostly by their observed relative offsets. The strict threshold value of 2.5 guarantees that the most perturbed sources do not get accepted for the input precious sets, and the derived systematic VSH fits of the ICRF3-Gaia fields are free of the astrophysical noise to the greatest possible degree. The downside of this approach is that a small fraction of sources with borderline physical offsets propagate into the precious set due to stochastic observational dispersion, while some unperturbed sources with the largest sky-correlated offsets may be rejected. Effectively, this is equivalent to using a weak Bayesian prior biasing the resulting VSH fits to smaller magnitudes. This problem has been dealt with by iterating the selection and VSH fitting procedures. We find that the updates to the working samples and fitting VSH coefficients are quite significant both in terms of the absolute values and S/N ratios. The iterations enhanced the VSH pattern of sky-correlated position differences by approximately a factor of 2, also achieving a higher normalized robustness of the fit. The greatest changes occurred for the smallest Ka sample, where the fitted field exceeds the formal coordinate errors. Our approach is also based on the assumption that the astrophysical radio-optical offsets are not sky-correlated. 
\citet{2022ApJ...939L..32S} predicted that non-blazar sources not currently exhibiting significant optical-radio offsets, which constitute the majority of ICRF3, are more likely to develop offsets in the future, further underscoring the need for continuous monitoring.

A maximally accurate, stable, and inertial multi-frequency reference frame has numerous applications. The sources in RORC-1 have the highest available formal precision and are likely to provide the greatest accuracy as well. They can be used for calibration purposes in geodetic VLBI measurements to achieve superior Earth orientation parameters and precision time determinations. RORC-1 sources are preferred for the important task of anchoring the Gaia CRF to ICRF3 in terms of global rotation. In this capacity, RORC sources can be preferentially observed and monitored by continuous VLBI sessions and form the core of ICRF4. They can be used to pin down systematics in VLBI-based CRFs such as the new Ka-band CRF, or, more generally, validate intermediate global solutions.

\begin{acknowledgements}
We thank the anonymous referee for a constructive and substantial review, which helped to improve this paper.
\end{acknowledgements}

\bibliography{mainbib}
\bibliographystyle{aasjournal}

\end{document}